\documentclass[aps, prapplied, amsmath, amssymb, floatfix, reprint, showkeys]{revtex4-2}
\usepackage[text={20cm,23cm},top=3cm]{geometry}  % This line determines the page layout.
\usepackage{amsmath,amsfonts,latexsym,mathtools,mathrsfs}
\usepackage{graphicx}
\usepackage{subfigure}
\usepackage[hidelinks]{hyperref}
\usepackage{xcolor}
\usepackage{titlesec}
\usepackage[titletoc,toc,title]{appendix}

\graphicspath{{Figures/}}

\numberwithin{equation}{section}

\newcommand{\lb}{\left (}
\newcommand{\rb}{\right )}

\newcommand{\lsq}{\left [}
\newcommand{\rsq}{\right ]}

\newcommand{\sechn}[2]{\mathrm{sech}^{#1} \lb #2 \rb}
\newcommand{\tanhn}[2]{\mathrm{tanh}^{#1} \lb #2 \rb}

\newcommand{\dd}[1]{\; \mathrm{d} #1}
\newcommand{\diff}[2]{\frac{\mathrm{d} #1}{\mathrm{d} #2}}
\newcommand{\diffn}[3]{\dfrac{\mathrm{d}^{#1} #2}{\mathrm{d} #3^{#1}}}

\renewcommand{\O}[1]{\mathcal{O} \lb #1 \rb}

\begin{document}
\title{Delamination Detection in Layered Waveguides using Ostrovsky Wave Packets}
\author{J. S. Tamber}
\author{D. J. Chappell}
\author{M. R. Tranter}\email{Matt.Tranter@ntu.ac.uk}\affiliation{
Department of Physics and Mathematics, Nottingham Trent University, NG11 8NS, United Kingdom}

\date{\today}

\begin{abstract}
We examine the scattering of Ostrovsky wave packets, generated from an incident solitary wave, in a two layered waveguide with a delamination in the centre and soft (imperfect) bonding either side of the centre.  The layers of the waveguide are assumed to consist of different materials, and the strains are described by a system of coupled Boussinesq equations. A semi-analytical approach consisting of matched asymptotic multiple-scale expansions is applied, leading to Ostrovsky equations in soft bonded regions and Korteweg-de Vries equations in the delaminated region. This semi-analytical method has good agreement with direct numerical simulations, validating the approach.

In the delaminated regions, Ostrovsky wave packets evolve into a train of solitary waves, which subsequently evolve into Ostrovsky wave packets in the second bonded region. Analysis of the phase shift in the wave packet, introduced from the delaminated region, allows us to predict both the position and the length of the delamination; the first time this has been achieved using nonlinear waves.  These results motivate experiments to validate the theoretical results, with the aim of creating a tool to monitor the integrity of layered structures.
\end{abstract}


\maketitle

\section{INTRODUCTION}
Extensive research has been undertaken on developing nonlinear models for strains in elastic waveguides, and many studies found that Boussinesq-type equations describe long longitudinal bulk strain waves in elastic waveguides, such as rods, bars and metal plates (see e.g. \cite{Samsonov01, Porubov03, Peake06, Khusnutdinova08, Peets17, Andrianov19, Garbuzov19}). Laboratory experiments have confirmed the existence of longitudinal bulk strain solitons in rods and bars \cite{Dreiden08, Dreiden11, Dreiden12, Dreiden14}. More recently, dispersive shock waves have been experimentally observed and theoretically studied for a viscoelastic waveguide \cite{Hooper21}. These experiments are focused on transparent materials to allow the use of interferometry for detecting strain waves.  We note also that linear waves methods can be employed in a wider range of materials at present, for example probing shock wave distribution in opaque solids \cite{Ducuosso21}.

In practical applications using layered structures,  it is imperative that the bonding between layers remains intact,  otherwise they could suffer catastrophic failure under stress. If we assume there is a flaw in the bonding such as delamination (where the material is completely debonded), the mathematical model takes the form of a scattering problem with continuity of longitudinal displacement and normal stress on the interface between bonded and delaminated sections. For a perfectly bonded waveguide (represented in experiments by cyanoacrylate), the displacements are described by Boussinesq equations \cite{Khusnutdinova08} and solitons fission in the delaminated regions providing a clear sign of delamination \cite{Khusnutdinova08, Khusnutdinova15}. For an imperfect `soft' bonding (represented in experiments by polychloroprene), a model based upon a series of anharmonic coupled dipoles gives rise to coupled regularised Boussinesq (cRB) equations describing displacements in a bi-layer, assuming sufficiently soft bonding \cite{Khusnutdinova09}. When the materials in the layers have similar properties, an incident solitary wave evolves in the bonded region into a solitary wave with a one-sided, co-propagating oscillatory tail, known as a radiating solitary wave. In the delaminated regions, the solitary wave detaches from its tail and this can be used to provide a measure of the delamination \cite{Khusnutdinova17}.  

In this paper, we focus on the case when the materials in each layer have significantly different properties. A limiting case was previously studied in \cite{Tamber22} where the displacement in the lower layer was assumed to be zero. The displacements are governed by Boussinesq-Klein-Gordon (BKG) equations with the leading-order strains described by Ostrovsky equations \cite{Ostrovsky78}. The evolution of wave packets generated from an initial pulse has been extensively studied \cite{Grimshaw08, Helfrich12, Grimshaw20}. In these studies, changes in the wave amplitude and phase shift in the transmitted wave field were used to identify the delamination length \cite{Khusnutdinova17, Tamber22, Tamber24}. Here we establish, for the first time, estimates for both the position and size of the delamination using theoretical properties of the Ostrovsky wave packet, a key development in the field. 

The paper is organised as follows. In Section \ref{sec:Problem} we outline the problem formulation for the scattering of zero mass wave packets in a bi-layer with a `sandwiched' delamination. An overview of the constructed weakly-nonlinear solution to the problem is presented in Section \ref{sec:WNL}. In Section \ref{sec:NumericalResults}, the direct numerical solution to the problem is compared to the constructed semi-analytical solution to confirm their agreement. Using observed changes in the transmitted wave field, we create `fans' consisting of the phase shift of the leading peak against delamination length, for a range of delamination positions. Using the data from both layers, or considering waves transmitted in both directions through the waveguide, we solve the inverse problem to uniquely determine both the position and length of delamination, laying the groundwork for experimental validation and the development of a technology for industry. 

% Problem formulation
\section{PROBLEM FORMULATION}
\label{sec:Problem}
\begin{figure}[h]
	\centering
	  \subfigure[Semi-infinite delamination case]{\includegraphics[width=6.4cm]{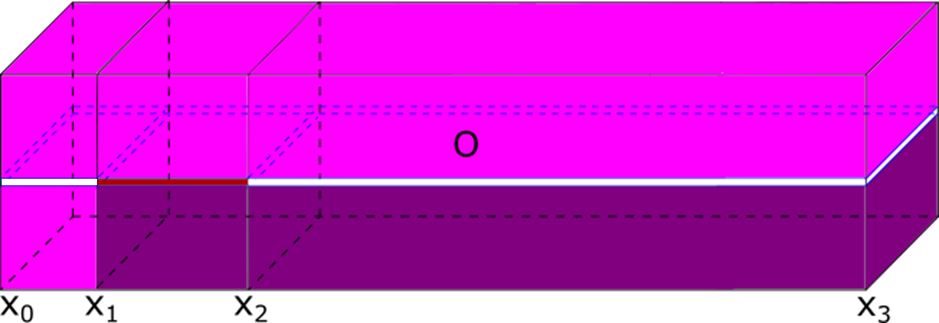}}
        \subfigure[Finite delamination case]{\includegraphics[width=6.4cm]{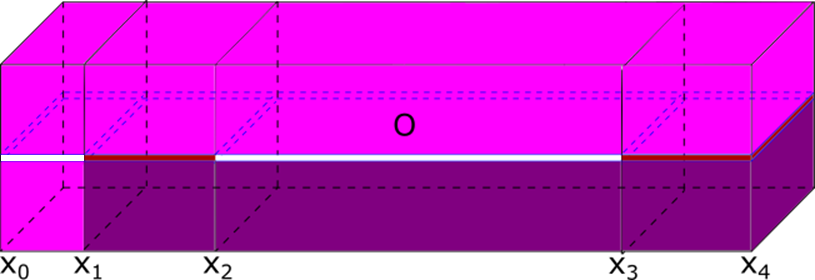}} 
	\caption{Schematic of two bi-layer structures that both include a homogeneous region for $x_0 < x < x_1$, a soft bonded region for $x_1 < x < x_2$ and a delaminated region for $x_2 < x < x_3$. In subplot (b) there is an additional soft bonded region for $x_3 < x < x_4$. We assume the homogeneous section has the same material properties throughout and for the other sections the materials in both layers have significantly different properties.}
	\label{fig:LayeredBar}
\end{figure}
Consider the layered waveguide illustrated in Figure \ref{fig:LayeredBar}, where we have two cases: a semi-infinite delamination that extends to the end of the waveguide, and a delamination sandwiched between two bonded sections. The inclusion of a homogeneous region is inspired by the experimental setup in \cite{Dreiden12}.  The limiting case when the material in the lower layer is not displaced by the initial impact is considered in \cite{Tamber22}, arising, for example,  when the lower layer has a much larger density than the upper layer.  In this paper we assume that the characteristic speeds in the layers are significantly different, but not the limiting case. The long longitudinal displacements in this system are governed by coupled regularised Boussinesq (cRB) equations in the bonded sections of the bar and uncoupled Boussinesq equations in the delaminated sections of the bar \cite{Khusnutdinova09, Khusnutdinova17}.

We denote longitudinal displacements in the upper layer for the $i$-th section as $u^{(i)}$ and for the lower layer as $w^{(i)}$. We are describing strain waves and therefore we set $f^{(i)} = u_{x}^{(i)}$ and $g^{(i)} = w_{x}^{(i)}$. For the soft bonded regions we have
\begin{align}
f_{tt}^{(i)} - f_{xx}^{(i)} &= 2 \epsilon \lsq -3 \lb {f^{(i)}}^2 \rb_{xx} + f_{ttxx}^{(i)} - \delta \lb f^{(i)} - g^{(i)} \rb \rsq, \notag \\
g_{tt}^{(i)} - c^2 g_{xx}^{(i)} &= 2 \epsilon \lsq -3 \alpha \lb {g^{(i)}}^2 \rb_{xx} + \beta g_{ttxx}^{(i)} + \gamma \lb f^{(i)} - g^{(i)} \rb  \rsq, \label{cRBS4Diff} 
\end{align}
for $i\in\{2,4\}$. The values of the constants $\alpha$, $\beta$ and $c$ depend on the physical and geometrical properties of the waveguide, while the constants $\delta$ and $\gamma$ depend on the properties of the soft bonding layer, and $\epsilon$ is a small amplitude parameter \cite{Khusnutdinova09, Khusnutdinova17}. For the homogeneous and delaminated regions we have the system
\begin{align}
f_{tt}^{(i)} - f_{xx}^{(i)} &= 2 \epsilon \lsq -3 \lb {f^{(i)}}^2 \rb_{xx} + f_{ttxx}^{(i)} \rsq, \notag \\
g_{tt}^{(i)} - c^2 g_{xx}^{(i)} &= 2 \epsilon \lsq -3 \alpha \lb {g^{(i)}}^2 \rb_{xx} + \beta g_{ttxx}^{(i)} \rsq \label{cRBS3Diff}
\end{align}
for $i\in\{1,3\}$, where $i=1$ represents the homogeneous region with $\alpha = \beta = c = 1$ and $i=3$ the delaminated region. 

These equations are complemented with continuity conditions at the interfaces between the sections, namely continuity of longitudinal displacement 
\begin{align}
u^{(i)} |_{x=x_i} = u^{(i+1)} |_{x=x_i}, \quad w^{(i)} |_{x=x_i} = w^{(i+1)} |_{x=x_i}, \quad i=1,2,3 \label{Cont}
\end{align}
and continuity of normal stress. The latter condition is imposed by writing the cRB equations \eqref{cRBS4Diff} in the form
\begin{align}
	f_{tt}^{(i)} &= \diffn{2}{\sigma_f^{(i)}}{x} - 2 \varepsilon \delta \lb f^{(i)} - g^{(i)} \rb, \notag \\
	g_{tt}^{(i)} &= \diffn{2}{\sigma_g^{(i)}}{x} + 2 \varepsilon \gamma \lb f^{(i)} - g^{(i)} \rb, \label{cRBSigma} 
\end{align}
so that we have
\begin{align}
\left. \sigma_f^{(i)} \right |_{x=x_i} &= \left. \sigma_f^{(i+1)} \right |_{x=x_i}, \label{Cont2_a} \\
\left. \sigma_g^{(i)} \right |_{x=x_i} &= \left. \sigma_g^{(i+1)} \right |_{x=x_i},\label{Cont2_b}
\end{align}
for $i\in\{1,3\}$. Note that the corresponding equations for $i\in\{2,4\}$ are obtained using (\ref{cRBS3Diff}) in place of (\ref{cRBS4Diff}), which results in a system of the form (\ref{cRBSigma}) with $\delta=\gamma=0$. 
We will investigate the scenario where the materials have significantly different properties, characterised by $c - 1 = \O{1}$.

\section{WEAKLY-NONLINEAR SOLUTION}
\label{sec:WNL}
To find an approximate solution to the system describing long longitudinal strains, we construct a weakly-nonlinear solution up to $\O{\epsilon}$ and use the continuity conditions to obtain `initial conditions' for the derived equations. This is implemented for the four-section waveguide here, with a sandwiched delamination, while for the three-section waveguide with semi-infinite delamination, the final soft bonded region can be discounted.

% First section - Homogeneous
\subsection{Homogeneous section}
\label{sec:Homogeneous Section}
Consider the first section of the bar shown in Figure 1(b), which is a homogenous region where the two layers are identical and thus the longitudinal displacements are governed by a Boussinesq equation in each layer. Therefore, following the notation from Section II, we introduce a leading order weakly-nonlinear solution in the first region of the structure,  which is described by (II.2), and takes the form
\begin{align*}
	f^{(1)} & \lb \xi, \eta, X \rb = g^{(1)}\lb \xi, \eta, X \rb\\
	&= I^{(1)} \lb \xi, X \rb + R^{(1)} \lb \eta, X \rb + \epsilon P^{(1)} \lb \xi, \eta, X \rb + \O{\epsilon^2},
\end{align*}
where we have introduced the characteristic variables $\xi = x - t$, $\eta = x + t$ and $X = \varepsilon x$.  The function $I^{(1)}$ represents the incident wave in the bar for both layers, which becomes a transmitted wave in other sections of the waveguide. Reflected waves are denoted by $R^{(1)}$ , and the higher order corrections at $\O{\epsilon}$ are given by $P^{(1)}$. Substituting this expansion for terms up to $\O{\epsilon}$ into the system (II.2) gives
\begin{align}
	-4P^{(1)}_{\xi \eta} &= \lb 2 I^{(1)}_{\xi X} -12 \left( I^{(1)}I^{(1)}_\xi\right)_\xi + 2 I^{(1)}_{\xi\xi\xi\xi} \rb  \notag \\
	&~~~+ \lb 2 R^{(1)}_{\eta X} -12 \left(R^{(1)} R^{(1)}_\eta\right)_\eta + 2 R^{(1)}_{\eta\eta\eta\eta}\rb \notag \\
	&~~~- 12 \lb I^{(1)}_{\xi\xi}R^{(1)} +  I^{(1)}R^{(1)}_{\eta\eta} + 2 I^{(1)}_\xi R^{(1)}_\eta \rb. \label{P1diff}
\end{align}
We can introduce fast space averaging for all sections of the bar in the form
\begin{equation}
    \lim_{x_{i-1} \rightarrow{-\infty}} \frac{1}{x_{i}-x_{i-1}}\int_{x_{i-1}}^{x_{i}} ... \mathrm{d}x
    \label{Averaging}
\end{equation}
for $i=1,2,3,4$. For $i=1$ here, the aim is to average at constant $\xi$ or $\eta$ to obtain equations in only $I^{(1)}$ or $R^{(1)}$, respectively. Averaging the left-hand side of \eqref{P1diff} in the fast space variable $x$ at constant $\eta$ gives
\begin{align}
    &~~~\lim_{x_0 \rightarrow{-\infty}} \frac{1}{x_{1}-x_0}\int_{x_0}^{x_{1}} P^{(1)}_{\xi \eta} \mathrm{d}x \notag \\
    &= \lim_{x_0 \rightarrow{-\infty}} \frac{1}{2(x_{1}-x_0)}\int_{2x_{0}-\eta}^{2x_{1}-\eta} P^{(1)}_{\xi \eta}\mathrm{d}\xi \notag \\
    &= \lim_{x_0 \rightarrow{-\infty}} \frac{1}{2(x_{1}-x_0)} \left[P^{(1)}_{\eta}\right]_{2x_{0}-\eta}^{2x_{1}-\eta} = 0.
\end{align}
A similar result can be found for constant $\xi$. Applying the averaging to the right-hand side yields two KdV equations of the form
\begin{align}
	I^{(1)}_{X} - 6 I^{(1)}I^{(1)}_\xi + I^{(1)}_{\xi\xi\xi} &= 0,
 \label{I1eq} \\
  R^{(1)}_{X} -6 R^{(1)} R^{(1)}_\eta + R^{(1)}_{\eta\eta\eta} &= 0. \label{R1eq}
\end{align}
Substituting \eqref{I1eq} and \eqref{R1eq} back into \eqref{P1diff}, then integrating twice, we obtain the higher order correction $P^{(1)}$ as
\begin{align}
	P^{(1)} &= 3 \lb 2 I^{(1)} R^{(1)} + R^{(1)}_{\eta} \int I^{(1)} \dd{\xi} + I^{(1)}_{\xi} \int R^{(1)} \dd{\eta} \rb \notag \\
	&~~~+ \phi^{(1)} \lb \xi, X \rb + \psi^{(1)} \lb \eta, X \rb,
	\label{P1}
\end{align}
where $\phi^{(1)}$, $\psi^{(1)}$ are arbitrary functions.

% Bonded region - uOst equations
\subsection{Imperfectly bonded sections}
\label{sec:Imperfectly Bonded Sections}
We now consider the case with soft bonding and we have $c - 1 = \O{1}$. The leading-order behaviour to the cRB equations should take the form of Ostrovsky wave packets \cite{Ostrovsky77} in agreement with the observed results in the corresponding initial-value problem \cite{Khusnutdinova11}. We present the derivation for both regions 2 and 4 of the waveguide here, but in the final region we assume that there is no leading order reflected wave (the waves have not reached the end of the waveguide and reflected back), so these terms can be omitted in region 4. 
Let us assume a weakly nonlinear solution of the form
\begin{align}
	f^{(i)} &= T^{(i)} \lb \xi, X \rb + R^{(i)} \lb \eta, X \rb + \epsilon P^{(i)} \lb \xi, \eta, X \rb + \O{\epsilon^2}, \label{f24}  \\
	g^{(i)} &= S^{(i)} \lb \nu, X \rb + G^{(i)} \lb \zeta, X \rb + \epsilon Q^{(i)} \lb \nu, \zeta, X \rb + \O{\epsilon^2},\label{g24}
\end{align}
$i=2,4$. The characteristic variables $\xi$, $\eta$ and $X$ are the same as before, but we now have different characteristic variables for the second layer, $\nu = x - ct$ and $\zeta = x + ct$. Substituting \eqref{f24} into (II.1) gives
\begin{align}
	-4P^{(i)}_{\xi \eta}  &= \left(2 T^{(i)}_{X} - 12 T^{(i)} T^{(i)}_\xi + 2 T^{(i)}_{\xi\xi\xi}\right)_{\xi} - \delta \left(T^{(i)}-S^{(i)}\right) \notag\\ 
&~~~ + \left(2 R^{(i)}_{X} -12 R^{(i)} R^{(i)}_\eta + 2R^{(i)}_{\eta\eta\eta}\right)_{\eta} -\delta \left(R^{(i)}-G^{(i)}\right) \notag\\
&~~~-12 \left(T^{(i)}_{\xi\xi}R^{(i)} + T^{(i)}R^{(i)}_{\eta\eta} + 2 T^{(i)}_\xi R^{(i)}_\eta\right), \label{P24diff}
\end{align}
for the upper layer, and substituting \eqref{g24} into (II.1) gives
\begin{align}
	-4c^2 Q^{(i)}_{\xi \eta}  &= \left(2c^2 S^{(i)}_{X} - 12 \alpha S^{(i)} S^{(i)}_\xi + 2 \beta c^2 S^{(i)}_{\xi\xi\xi}\right)_{\xi} \notag \\
	&~~~+ \gamma \left(T^{(i)}-S^{(i)}\right) + \gamma \left(R^{(i)}-G^{(i)}\right) \notag\\
&~~~ + \left(2c^2 G^{(i)}_{X} - 12 \alpha G^{(i)} G^{(i)}_\eta + 2 \beta c^2 G^{(i)}_{\eta\eta\eta}\right)_{\eta} \notag \\
&~~~-12 \alpha \left(S^{(i)}_{\xi\xi}G^{(i)} + S^{(i)}G^{(i)}_{\eta\eta} + 2 S^{(i)}_\xi G^{(i)}_\eta\right), \label{Q24diff}
\end{align}
for the lower layer. Applying the averaging \eqref{Averaging} for $i=2,4$, holding each of the characteristic variables (except $X$) constant, leads to four uncoupled Ostrovsky equations of the form
\begin{align}
    \left( R^{(i)}_{X} - 6 R^{(i)} R^{(i)}_\eta + R^{(i)}_{\eta\eta\eta} \right)_{\eta} &= \delta R^{(i)}, \label{R24} \\
    \left( T^{(i)}_{X} - 6 T^{(i)} T^{(i)}_\xi + T^{(i)}_{\xi\xi\xi} \right)_{\xi} &= \delta T^{(i)}, \label{T24} \\
    \left( c^2 G^{(i)}_{X} - 6 \alpha G^{(i)} G^{(i)}_{\zeta} + \beta c^2 G^{(i)}_{\zeta\zeta\zeta} \right)_{\zeta} &= \gamma G^{(i)}, \label{G24} \\
    \left( S^{(i)}_{X} - 6 \alpha S^{(i)} S^{(i)}_{\nu} + \beta c^2 S^{(i)}_{\nu\nu\nu} \right)_{\nu} &= \gamma S^{(i)}. \label{S24}
\end{align}
Substituting these back into \eqref{P24diff} and \eqref{Q24diff} we find
\begin{align}
	-4P^{(i)}_{\xi \eta} &= -12 \left(T^{(i)}_{\xi\xi}R^{(i)} + T^{(i)}R^{(i)}_{\eta\eta} + 2 T^{(i)}_\xi R^{(i)}_\eta\right) \notag \\
	&~~~+ \delta G^{(i)} + \delta S^{(i)}, \label{P24diff2} \\
	-4c^2 Q^{(i)}_{\xi \eta}  &= -12 \alpha \left(S^{(i)}_{\xi\xi}G^{(i)} + S^{(i)}G^{(i)}_{\eta\eta} + 2 S^{(i)}_\xi G^{(i)}_\eta\right) \notag \\
	&~~~+ \gamma T^{(i)} + \gamma S^{(i)}. \label{Q24diff2}
\end{align}
In \eqref{P24diff2} and \eqref{Q24diff2} we have terms involving the other characteristic variables, namely $S^{(i)}$ and $G^{(i)}$ for \eqref{P24diff2}, with $T^{(i)}$ and $R^{(i)}$ appearing in \eqref{Q24diff2}. We can rewrite these in terms of the corresponding characteristic variables for $P^{(i)}$ and $Q^{(i)}$, to allow us to perform the integration. For example, we can rewrite $\zeta = x - ct$ as 
\begin{equation}
    \zeta = \frac{(1-c)\xi+(1+c)\eta}{2}. 
    	\label{zeta}
\end{equation}
Making the appropriate change for each of the terms, we can then integrate to find expressions for $P^{(i)}$ and $Q^{(i)}$ as
\begin{align}
	P^{(i)} &= \hat{G} + \hat{S} + h_1 + \phi_1^{(i)} \lb \xi, X \rb + \psi_1^{(i)} \lb \eta, X \rb, \label{P24eq} \\
	Q^{(i)} &= \hat{R} + \hat{T} + h_2 + \phi_2^{(i)} \lb \nu, X \rb + \psi_2^{(i)} \lb \zeta, X \rb, \label{Q24eq}
\end{align}
where $\phi_{1,2}^{(i)}$ and $\psi_{1,2}^{(i)}$ are arbitrary functions and we have introduced
\begin{align*}
	\hat{G} &= -\frac{\delta}{1-c^2}\int_{-\infty}^{\zeta} \int_{-\infty}^{v} G^{(i)}(u, X) \mathrm{d}u \mathrm{ d}v, \\
	\hat{S} &= -\frac{\delta}{1-c^2} \int_{-\infty}^{\nu} \int_{-\infty}^{v} S^{(i)}(u, X) \mathrm{d}u \mathrm{ d}v, \\
	\hat{R} &= \frac{\gamma}{1-c^2} \int_{-\infty}^{\eta} \int_{-\infty}^{v} R^{(i)}(u, X) \mathrm{d}u \mathrm{ d}v, \\
	\hat{T} &= \frac{\gamma}{1-c^2}  \int_{-\infty}^{\xi} \int_{-\infty}^{v} T^{(i)}(u, X) \mathrm{d}u \mathrm{ d}v, \\
  h_1 &= 3 \lb 2 T^{(i)} R^{(i)} + R^{(i)}_{\eta} \int T^{(i)} \dd{\xi} + T^{(i)}_{\xi} \int R^{(i)} \dd{\eta} \rb, \\
   h_2 &= 3 \alpha \lb 2 S^{(i)} G^{(i)} + G^{(i)}_{\eta} \int S^{(i)} \dd{\nu} + S^{(i)}_{\xi} \int G^{(i)} \dd{\zeta} \rb.
\end{align*}

\subsection{Delaminated section}
\label{sec:Delaminated Section}
In the third region of the waveguide, where the delamination is present we employ a similar methodology as described in Section \ref{sec:Homogeneous Section} but as the Boussinesq equations have different coefficients we need to use a different set of characteristic variables in each layer, as in Section \ref{sec:Imperfectly Bonded Sections}. Namely, we take a leading order weakly-nonlinear solution in the form
\begin{align*}
	f^{(3)} &= T^{(3)} \lb \xi, X \rb + R^{(3)} \lb \eta, X \rb + \epsilon P^{(3)} \lb \xi, \eta, X \rb + \O{\epsilon^2}, \\
	g^{(3)} &= S^{(3)} \lb \nu, X \rb + G^{(3)} \lb \zeta, X \rb + \epsilon Q^{(3)} \lb \nu, \zeta, X \rb + \O{\epsilon^2}.
\end{align*}
Substituting into (II.2) we obtain
\begin{align}
	-4P^{(3)}_{\xi \eta} &= \lb 2 I^{(3)}_{\xi X} -12 \left( I^{(3)}I^{(3)}_\xi\right)_\xi + 2 I^{(3)}_{\xi\xi\xi\xi} \rb  \notag \\
	&~~~+ \lb 2 R^{(3)}_{\eta X} -12 \left(R^{(3)} R^{(3)}_\eta\right)_\eta + 2 R^{(3)}_{\eta\eta\eta\eta}\rb \notag \\
	&~~~- 12 \lb I^{(3)}_{\xi\xi}R^{(3)} +  I^{(3)}R^{(3)}_{\eta\eta} + 2 I^{(3)}_\xi R^{(3)}_\eta \rb, \label{P3diff}
\end{align}
for the upper layer, and for the lower layer we find
\begin{align}
	-4c^2 Q^{(3)}_{\nu \zeta} &= 2c^2 S^{(3)}_{\nu X} - 12 \alpha \left( S^{(3)}S^{(3)}_\nu\right)_\nu + 2 \beta c^2 S^{(3)}_{\nu\nu\nu\nu}   \notag \\
	&~~~+ 2c^2 G^{(3)}_{\zeta X} - 12 \alpha \left(G^{(3)} G^{(3)}_\zeta\right)_\zeta + 2 \beta c^2 G^{(3)}_{\zeta\zeta\zeta\zeta}\ \notag \\
	&~~~- 12 \alpha \lb S^{(3)}_{\nu\nu} G^{(3)} +  S^{(3)} G^{(3)}_{\zeta\zeta} + 2 S^{(3)}_\nu G^{(3)}_\zeta \rb. \label{Q3diff}
\end{align}
Applying the averaging \eqref{Averaging}, holding each of the respective characteristic variables constant, leads to KdV equations for the transmitted waves of the form
\begin{align}
	T^{(3)}_{X} - 6 T^{(3)}T^{(3)}_\xi + T^{(3)}_{\xi\xi\xi} &= 0,
 \label{T3eq} \\
  	S^{(3)}_{X} - 6 \frac{\alpha}{c^2} S^{(3)} S^{(3)}_\nu + \beta S^{(3)}_{\nu\nu\nu} &= 0, \label{S3eq}
\end{align}
while for the reflected waves we find
\begin{align}
	R^{(3)}_{X} - 6 R^{(3)}R^{(3)}_\eta + R^{(3)}_{\eta\eta\eta} &= 0,
 \label{R3eq} \\
  	G^{(3)}_{X} - 6 \frac{\alpha}{c^2} G^{(3)} G^{(3)}_\zeta + \beta G^{(3)}_{\zeta\zeta\zeta} &= 0. \label{G3eq}
\end{align}
Substituting \eqref{T3eq} - \eqref{G3eq} into \eqref{P3diff} and \eqref{Q3diff}, then integrating twice, gives expressions for the higher-order corrections of the form
\begin{align}
	P^{(3)} &= 3 \lb 2 T^{(3)} R^{(3)} + R^{(3)}_{\eta} \int T^{(3)} \dd{\xi} + T^{(3)}_{\xi} \int R^{(3)} \dd{\eta} \rb \notag \\
	&~~~+ \phi_1^{(3)} \lb \xi, X \rb + \psi_1^{(3)} \lb \eta, X \rb,
	\label{P3}
\end{align}
and
\begin{align}
	Q^{(3)} &= 3 \alpha \lb 2 S^{(3)} G^{(3)} + G^{(3)}_{\zeta} \int S^{(3)} \dd{\nu} + S^{(3)}_{\nu} \int G^{(3)} \dd{\zeta} \rb \notag \\
	&~~~+ \phi_2^{(3)} \lb \nu, X \rb + \psi_2^{(3)} \lb \zeta, X \rb,
	\label{Q3}
\end{align}
where $\phi_{1,2}^{(3)}$ and $\psi_{1,2}^{(3)}$ are arbitrary functions.

% Matching equations at the boundaries
\subsection{Matching at the boundaries}
\label{sec:BC}
We use the continuity conditions \eqref{Cont},  \eqref{Cont2_a} and \eqref{Cont2_b} to establish `initial conditions' for the equations derived within each region of the waveguide. The weakly-nonlinear solution in each region is substituted into the appropriate continuity conditions, noting that we differentiate \eqref{Cont} with respect to time, as in previous works \cite{Khusnutdinova15, Khusnutdinova17, Tamber22}. We can then express the transmitted and reflected wave in terms of the wave incident on the boundary, giving rise to reflection and transmission coefficients.

For the upper layer, the equation system that we solve is identical to the one considered in \cite{Tamber22}, where we found
\begin{align}
	&R^{(i+1)}|_{x=x_i} = C_R^{(i)} T^{(i)}|_{x=x_i}, &&T^{(i+1)}|_{x=x_i} = C_T^{(i)} T^{(i)}|_{x=x_i},
	\label{RTICxa}
\end{align}
$i=1,2,3$.  As the material is the same throughout the upper layer, we find the reflection and transmission coefficients are $C_R^{(i)} = 0$ and $C_T^{(i)} = 1$, respectively, representing full transmission of the incident wave at leading order. 

For the lower layer, the equation system is similar, namely 
\begin{align}
	&G^{(i+1)}|_{x = x_i} = C_G^{(i+1)} S^{(i)}|_{x = x_i}, &&S^{(i+1)}|_{x = x_i} = C_S^{(i+1)} S^{(i)}|_{x = x_i},
	\label{GSICxa}
\end{align}
$i=1,2,3$. In regions 2 - 4, the materials in the lower layer are the same in each region and so we have full transmission as with the upper layer. For the interface between the first and second region, our transmission and reflection coefficients take the form
\begin{align}
	C_G^{(1)} = \frac{c - 1}{c + 1}, \quad C_S^{(1)} = \frac{2}{c \lb 1 + c \rb}. 
	\label{CRCTxa}
\end{align}

%Numerical Results
\section{NUMERICAL RESULTS}
\label{sec:NumericalResults}
We use two numerical techniques for solving the systems of equations derived here. For the system of equations \eqref{cRBS4Diff} - \eqref{cRBS3Diff}, in conjunction with the continuity conditions \eqref{Cont}, \eqref{Cont2_a} and \eqref{Cont2_b}, we perform direct numerical simulations based upon the scheme outlined in \cite{Tranter19}. For the weakly-nonlinear solution, we employ a semi-analytical method by solving the derived equations using a pseudospectral method, which aligns with the methodology utilised in \cite{Khusnutdinova17, Tamber22}.

For the direct numerical simulations we use step sizes of $\Delta x = \Delta t = 0.01$ and, for the pseudospectral method, we use a step size of $0.1$ for the characteristic variables $\xi,  \eta,  \nu, \zeta$ and take $\Delta X = 0.0005$. The coefficients are $\alpha = \beta = 1 + \varepsilon$, $\delta = \gamma = 0.5$, and $\epsilon = 0.005$. We assume that the displacement at the boundary where the incident wave enters is constant and that the wave propagates into an unperturbed medium, therefore we have boundary conditions $u_{x} = f = 0$ at both ends of the bar. This is the same boundary condition that was imposed for the development of the numerical method in \cite{Tranter19}.

The Ostrovsky equation clearly has a zero-mean constraint, found by integrating the equation in the spatial variable e.g.
\begin{equation*}
	\frac{1}{2L} \int_{-L}^{L} I(\xi, X) \dd{\xi}.
\end{equation*}
In the initial-value problem, this can be overcome by explicitly calculating the mean value of the Boussinesq-Klein-Gordon \cite{Khusnutdinova19} or cRB equations \cite{Khusnutdinova22} and constructing a weakly-nonlinear solution in the regime where all functions necessarily have zero-mean. This approach is yet to be adapted for the scattering problem, so we circumvent this issue by considering localised strains that already satisfy the zero mean requirement \cite{Khusnutdinova17, Tamber22}.

Explicitly, for the weakly-nonlinear solution, the initial condition is assumed to be the exact solitary wave solution to the incident wave equation \eqref{I1eq}, with an adjustment for zero-mean, namely
\begin{align}
	I \lb \xi, X_0 \rb &:= A\ \sechn{2}{\frac{\xi}{\Lambda}} \notag \\&~~~- \frac{\Gamma}{10} \lsq \sechn{2}{\frac{\xi + 5}{10 \Lambda}} + \sechn{2}{\frac{\xi - 5}{10 \Lambda}} \rsq,
	\label{IIC}	
\end{align}
where $A = -\frac{v}{2}$, $\Lambda = \frac{2}{\sqrt{v}}$ and 
\begin{equation*}
	\Gamma =  \frac{A\ \tanhn{}{\frac{L}{\Lambda}}}{\tanhn{}{\frac{L + 5}{10 \Lambda}} + \tanhn{}{\frac{L - 5}{10\Lambda}}}.
\end{equation*}
This gives a solitary wave with a long well to spread the mass. To determine the initial conditions in other sections of the bi-layer, we make use of the relations outlined in Section \ref{sec:BC} to express them in terms of \eqref{IIC}. The corresponding initial conditions for the cRB equations are chosen as
\begin{equation}
	u(x, 0) = w(x, 0) = I(x, 0), \quad u_t(x,0) = w_t(x, 0) = -\diff{I}{\xi}(x, 0),
\end{equation}
which ensures that we only have a right-propagating wave in regions 2 to 4 of the bar at leading order.

\subsection{Semi-infinite delamination}
\begin{figure*}[t] 
	\center
	\includegraphics[width = 12.9cm]{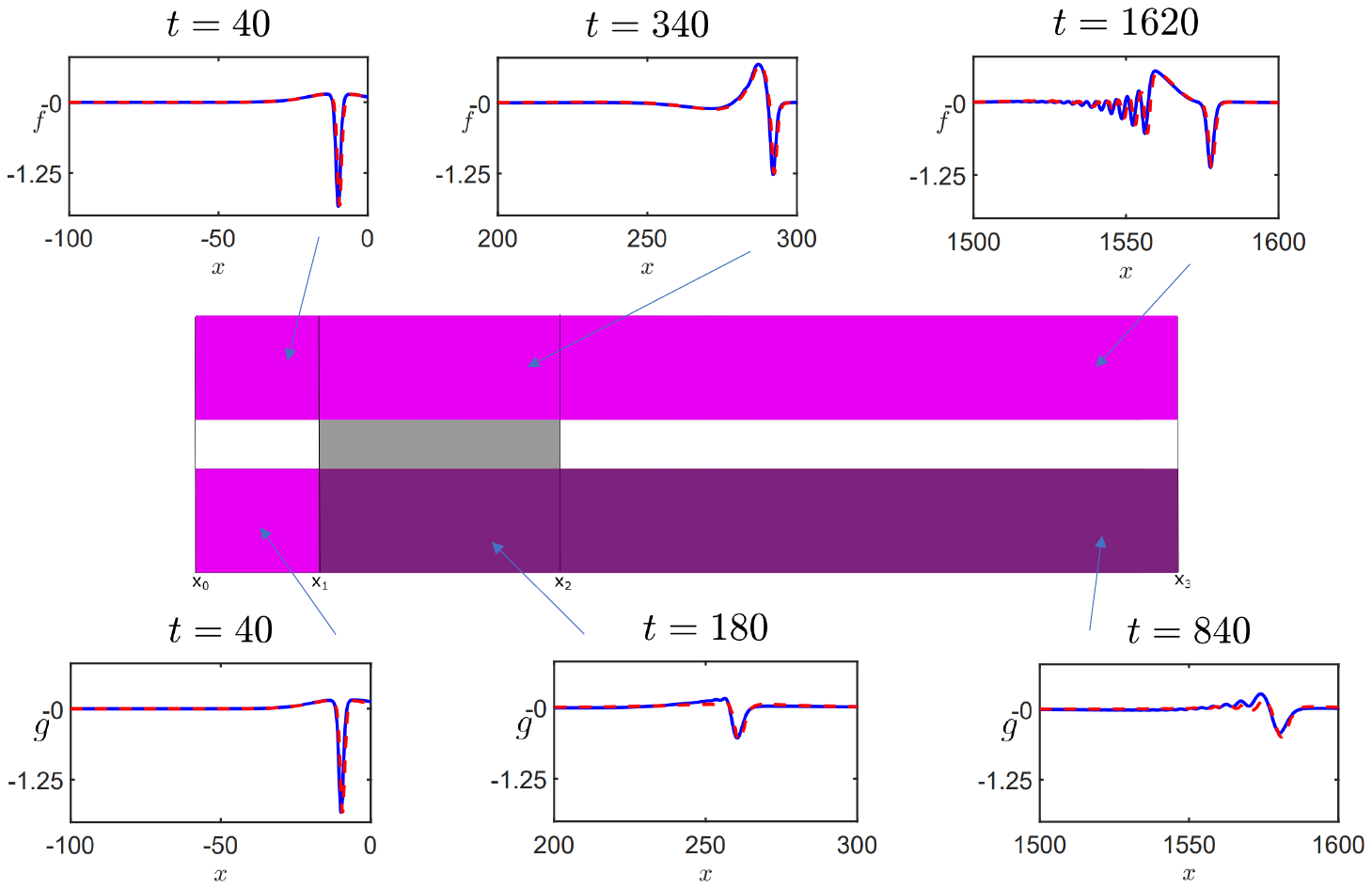}
	\caption{Waves in a bar with semi-infinite delamination computed via direct numerical simulations (blue, solid line) and the semi-analytical method (red, dashed line). Parameters: $v = 4$, $c = 1.5$, $\alpha = \beta = 1 + \varepsilon$, $\delta = \gamma = 0.5$. For the finite-difference method, the full computational domain is $[-100, 1600]$. In the pseudospectral method, we have $N = 131,072$.}
	\label{fig:SemiInfiniteComparison}
\end{figure*}

Firstly we examine the bi-layer waveguide featuring semi-infinite delamination, as depicted in Figure \ref{fig:LayeredBar}(a), and compare the outcomes of direct numerical simulations against semi-analytical results. The size of each section of the bar is specified by taking $x_0=-100$, $x_1=0$, $x_2=300$ and $x_3=1600$. The comparative analysis, as illustrated in Figure \ref{fig:SemiInfiniteComparison}, demonstrates an excellent agreement across all regions in the waveguide. In the upper layer, the incident soliton evolves into an Ostrovsky wave packet within the bonded region, consistent with the anticipated outcomes outlined in \cite{Khusnutdinova19, Tamber22}. This wave packet undergoes evolution into solitons within the delaminated region. We observe a lead soliton that has almost fully separated from the well to travel on a zero background, while others could still be trapped and would appear with a larger computational domain.

In the lower waveguide, the wave exhibits behavior similar to that observed in the upper waveguide. However, due to the higher wave speed and shorter duration spent in each region, we observe comparatively less evolution. The accuracy of the semi-analytical method could be improved by the inclusion of higher-order terms or by reducing the value of $\varepsilon$.

\subsection{Finite delamination}
\label{sec:FiniteDelam}
The next natural case to consider is when the delamination is sandwiched between two soft bonded regions, as depicted in Figure \ref{fig:LayeredBar}(b). We firstly aim to show agreement between the two numerical schemes, before considering how to detect delamination.

\begin{figure}[h!] 
    \centering
    \subfigure{\includegraphics[width=4.2cm]{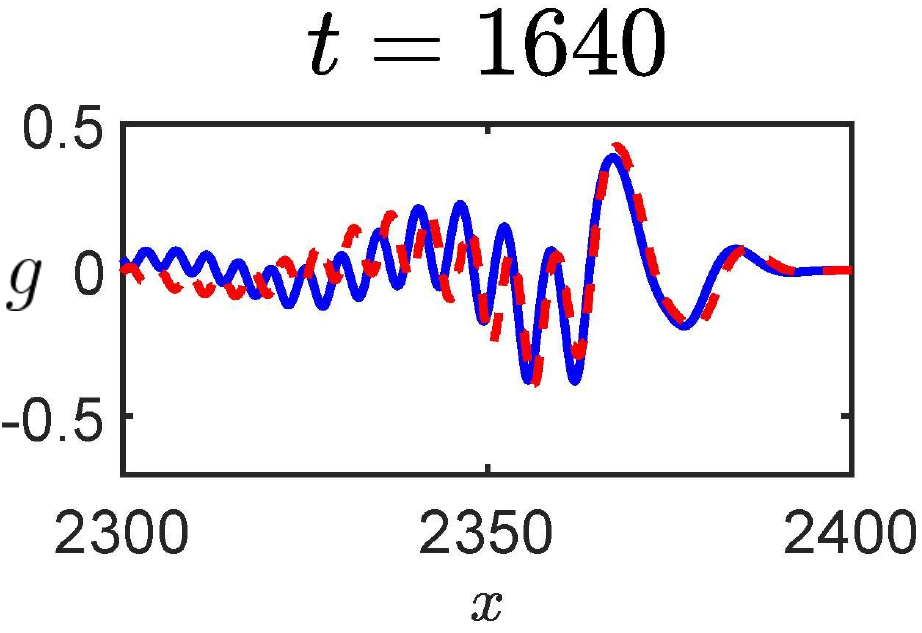}} \hspace{0.1cm}
    \subfigure{\includegraphics[width=4.2cm]{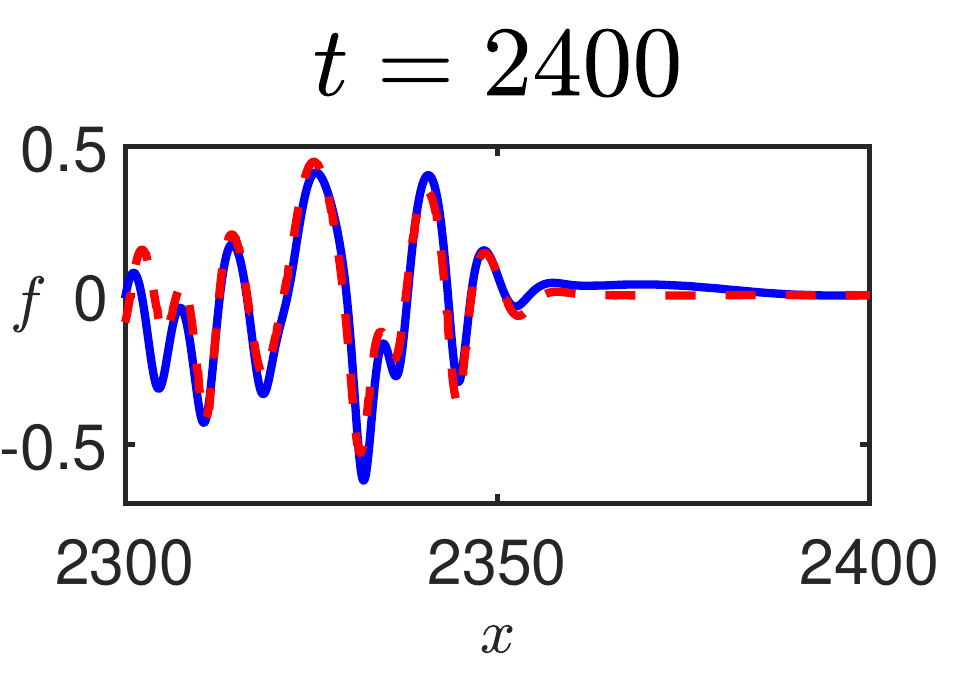}}
    \caption{Comparison of the solution schemes in the final section of the bar shown in Figure \ref{fig:LayeredBar}(b), at $t=1640$ for the lower waveguide and $t=2400$ for the upper waveguide. The parameters are the same as for Figure \ref{fig:SemiInfiniteComparison}, with the direct numerical simulation (blue, solid line) and semi-analytical solution (red, dashed line) in good agreement. For the finite-difference method, the full computational domain is $[-100, 2400]$. In the pseudospectral method, we have $N = 131,072$.}
    \label{fig:Finite}
\end{figure}

We modify the semi-infinite delamination scenario by introducing an additional soft bonded region for $1600 < x < 2400$. The results in the first three sections can be seen in Figure \ref{fig:SemiInfiniteComparison}, while the comparison in the final region is shown in Figure \ref{fig:Finite}. In both layers, we observe that any emerging solitary waves are now evolving into Ostrovsky wave packets, leading to a complex behaviour. In comparison to the previous soft bonded region there are significantly more peaks and the amplitude has also decreased significantly, a clear sign of delamination. It is worth noting that, while there is good agreement around the leading wave packet, the agreement weakens elsewhere due to the absence of higher-order correction terms.

\subsection{Characterising the delamination}
\label{sec:Characterising}
To understand how the length and position of the delaminated region impact a bonded structure, we must explore how varying the delamination position, represented by $x_2$, and the delamination length, denoted as $D = x_3 - x_2$, influence the waves within the final bonded region. We observed previously that the wave structure changes significantly, but we require a direct measure that can be used for classification.

\begin{figure}[h!] 
\centering
	\subfigure[$D=0$]{\includegraphics[width=4.3cm]{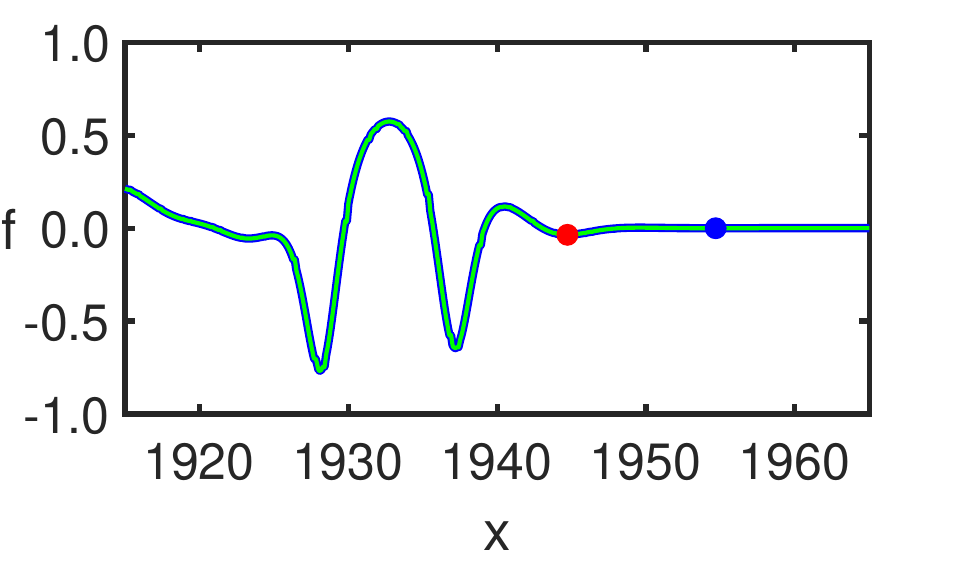}}~
	\subfigure[$D=500$]{\includegraphics[width=4.3cm]{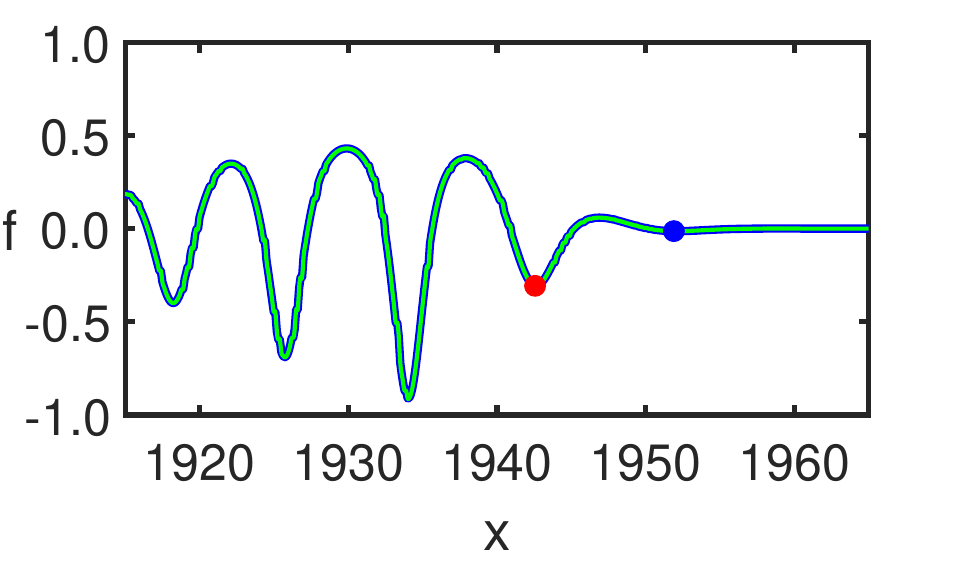}}\\
	\subfigure[$D=1000$]{\includegraphics[width=4.3cm]{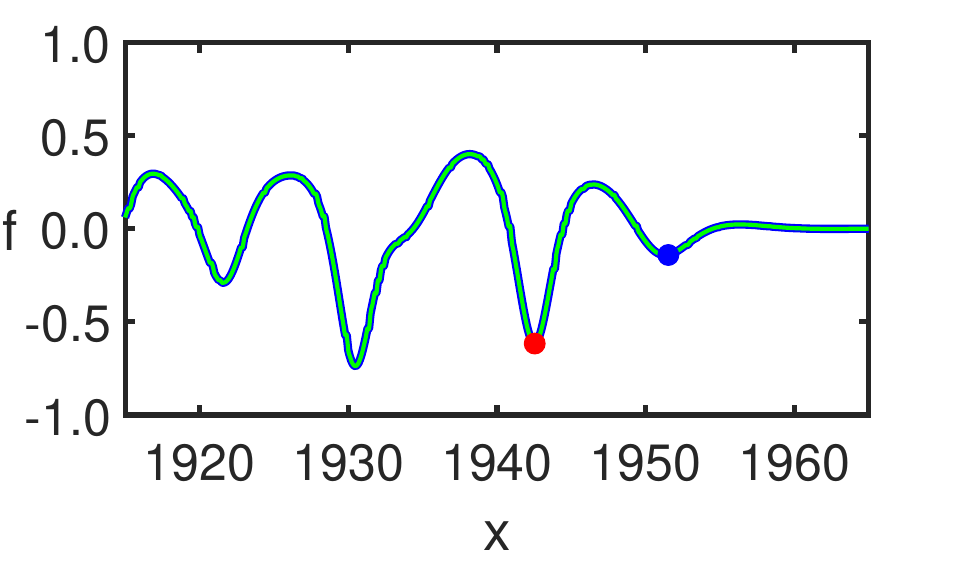}}~
	\subfigure[$D=1500$]{\includegraphics[width=4.3cm]{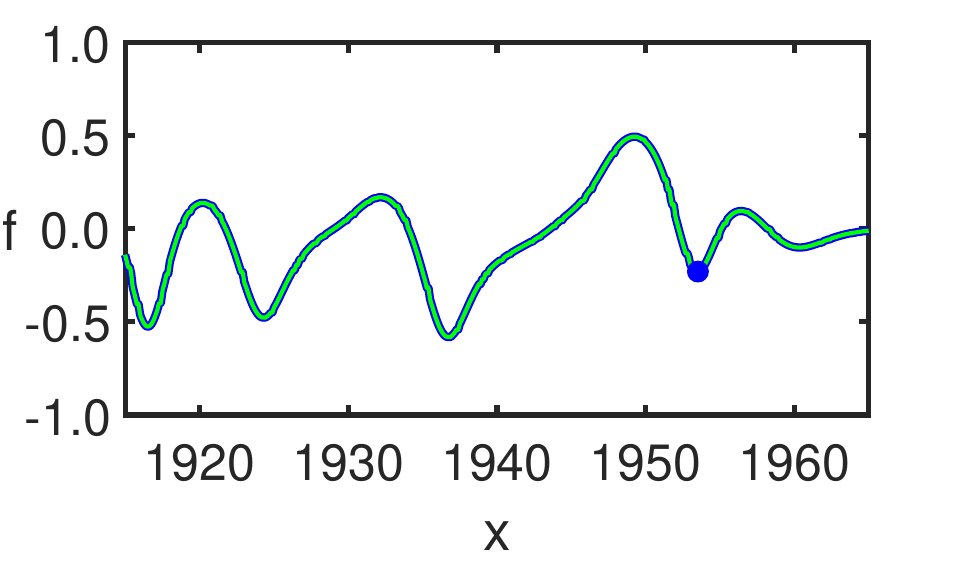}}
    	\caption{Identifying the leading peak in the Ostrovsky wave packet, represented by red and blue dots, for various delamination lengths.}
	\label{fig:NewLeadingPeak}
\end{figure}

In \cite{Tamber22} it was noted that the linear dispersion relation for the derived KdV or Ostrovsky equations can help determine changes in the wave packet. Indeed, the wave speed of the Ostrovsky wave packet is the maximum of the group speed ($< c$), while solitons move with a speed in the phase gap ($> c$). Therefore, for a longer delaminated region, a larger phase shift will be introduced. However, this can be masked by the complex evolution of the wave packet.  In order to identify a characteristic that can be used to characterise the delamination, we plot the solution for four different values of delamination length in Figure \ref{fig:NewLeadingPeak}.

In Figure \ref{fig:NewLeadingPeak}(a),  when there is no delamination, we can see a small leading peak in the wave packet (red dot) and a subsequent point of interest is also shown (blue dot). As the delamination length is increased, this leading peak increases in amplitude and begins to move backwards, while the second point of interest has begun to develop into a peak. Therefore, the overall wave packet has moved forward, despite the leading peak moving backwards. Finally, for large delaminations, the first peak is not identifiable in the main wave packet, but the second peak is clear (there is also a third peak forming). This shift in the overall wave packet agrees with the expectations of the linear dispersion relation analysis.

Manually measuring the front of the wave packet can be difficult, although this was done with some success in \cite{Tamber22}. Therefore, in the subsequent sections of this paper, we will employ spline interpolation to interpret the data between discretised points and improve the tracking of the leading peak.

\begin{figure}[h!] 
    \centering
    \subfigure{\includegraphics[width=4.7cm]{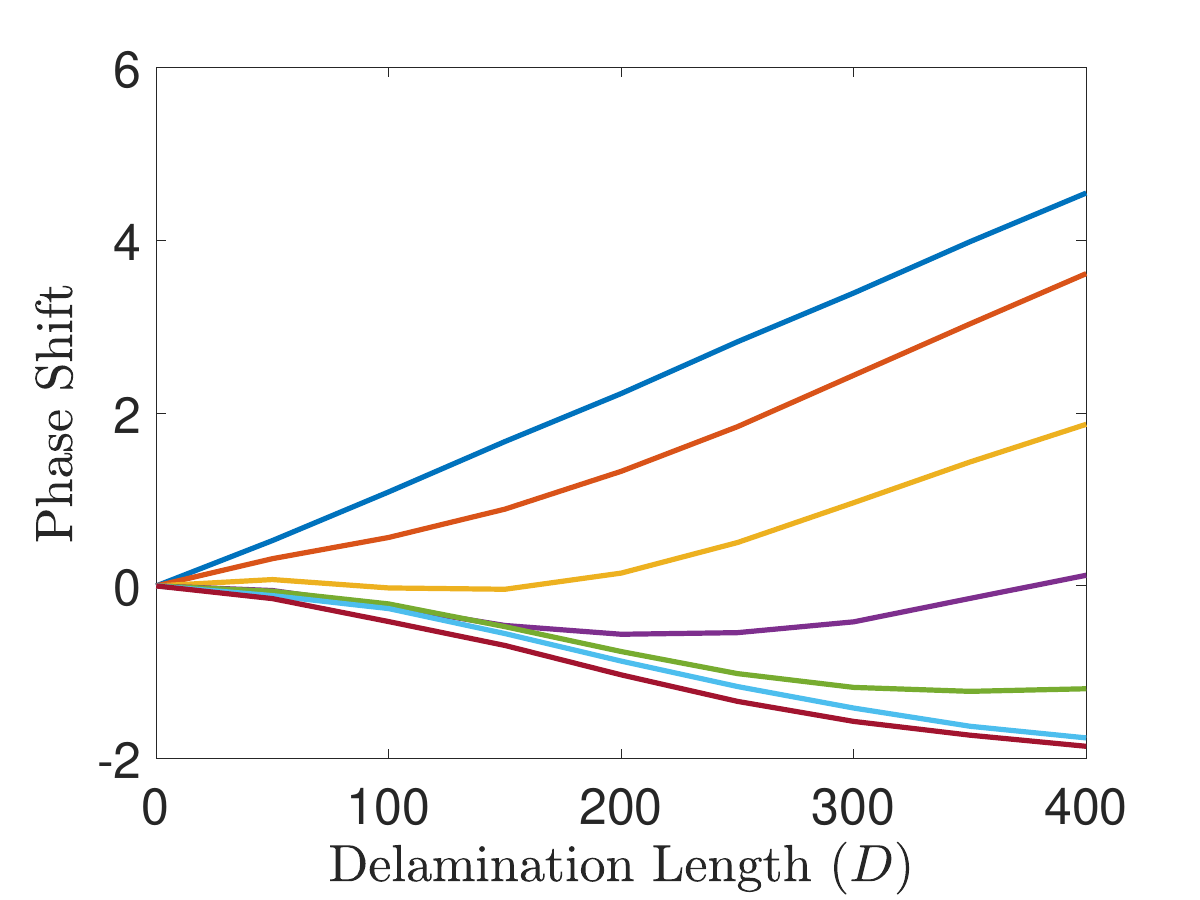}}
    \subfigure{\includegraphics[width=4.7cm]{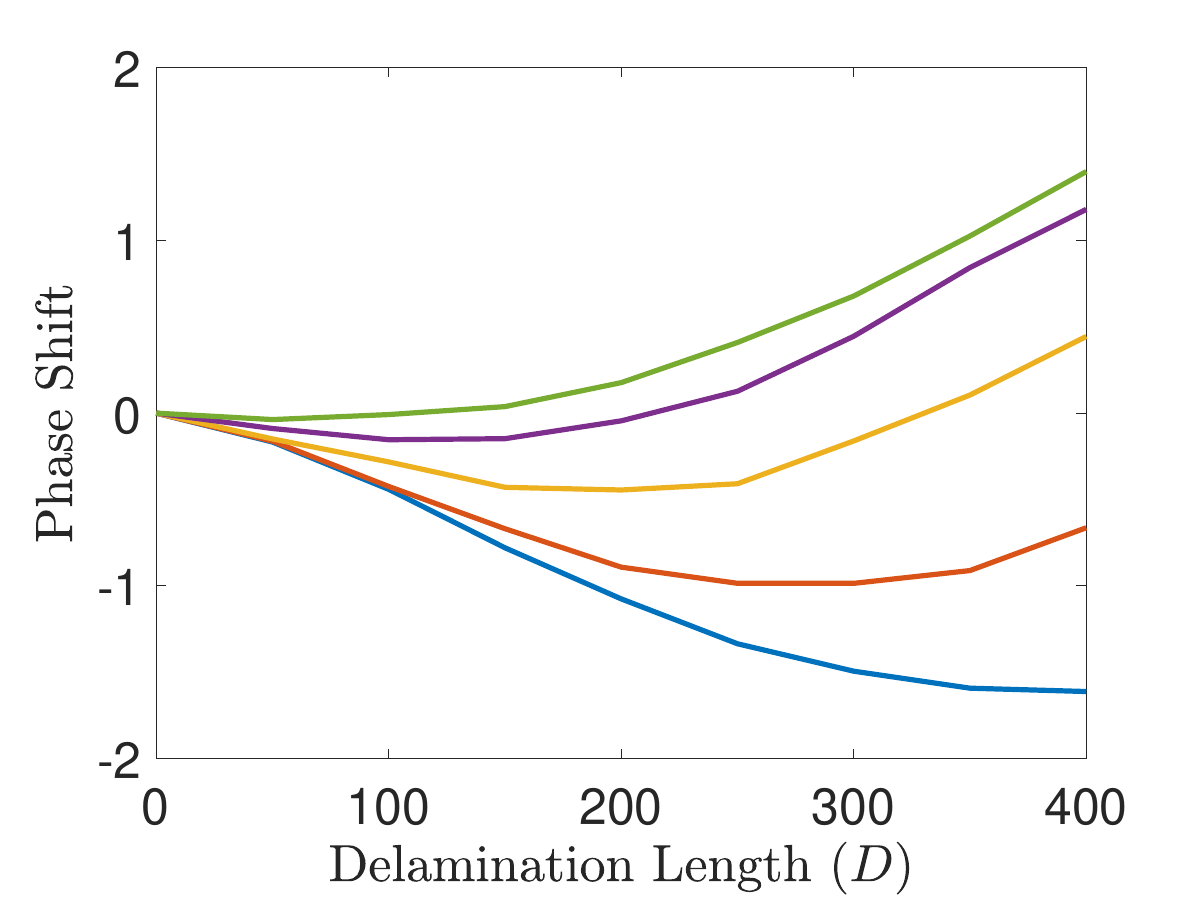}}\\
	\addtocounter{subfigure}{-2}
    \subfigure[{$x_2\in [0,300]$}]{\includegraphics[width=4.3cm]{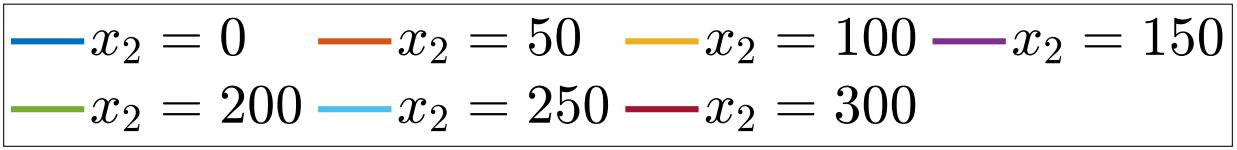}}
    \hspace{0.2cm}
    \subfigure[{$x_2\in [350, 550]$}]{\includegraphics[width=4.3cm]{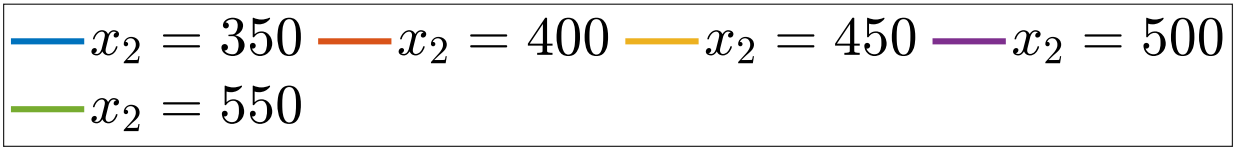}}
    \caption{`Fan' plots of the change in phase shift in the upper layer, with reference to the fully bonded case, compared against changes in delamination length. The position $x_2$ is varied from (a) 0 to 300 and (b) 350 to 550, and the delamination length $D$ is varied from 0 to 400.}
    \label{fig:FanUpper}
\end{figure}

Firstly, we focus on the upper layer within the structure, using the same parameters as Section \ref{sec:FiniteDelam}. The delamination length is varied from $D = 0$ in increments of 10 up to $D = 400$. Simultaneously, the delamination position is varied from $x_2 = 0$ to $x_2 = 550$. The maximum computational time considered for the calculation is set at $t = 2000$. By tracking the peak, we can measure the phase shift in comparison to the case without delamination. The result is plotted in Figure \ref{fig:FanUpper}.

For $x_2=0$ to $x_2=300$ the phase shift curve decreases in gradient, creating a `fan' shape, and then for $x_2=350$ to $x_2=550$ the phase shift curve is increasing in gradient. Hence, for an observed phase shift, we may obtain multiple possible values for the delamination length and position due to multiple intercepts with the curve, and for some values the curves may coincide. 

\begin{figure}[h!] 
    \centering
    \subfigure{\includegraphics[width=4.7cm]{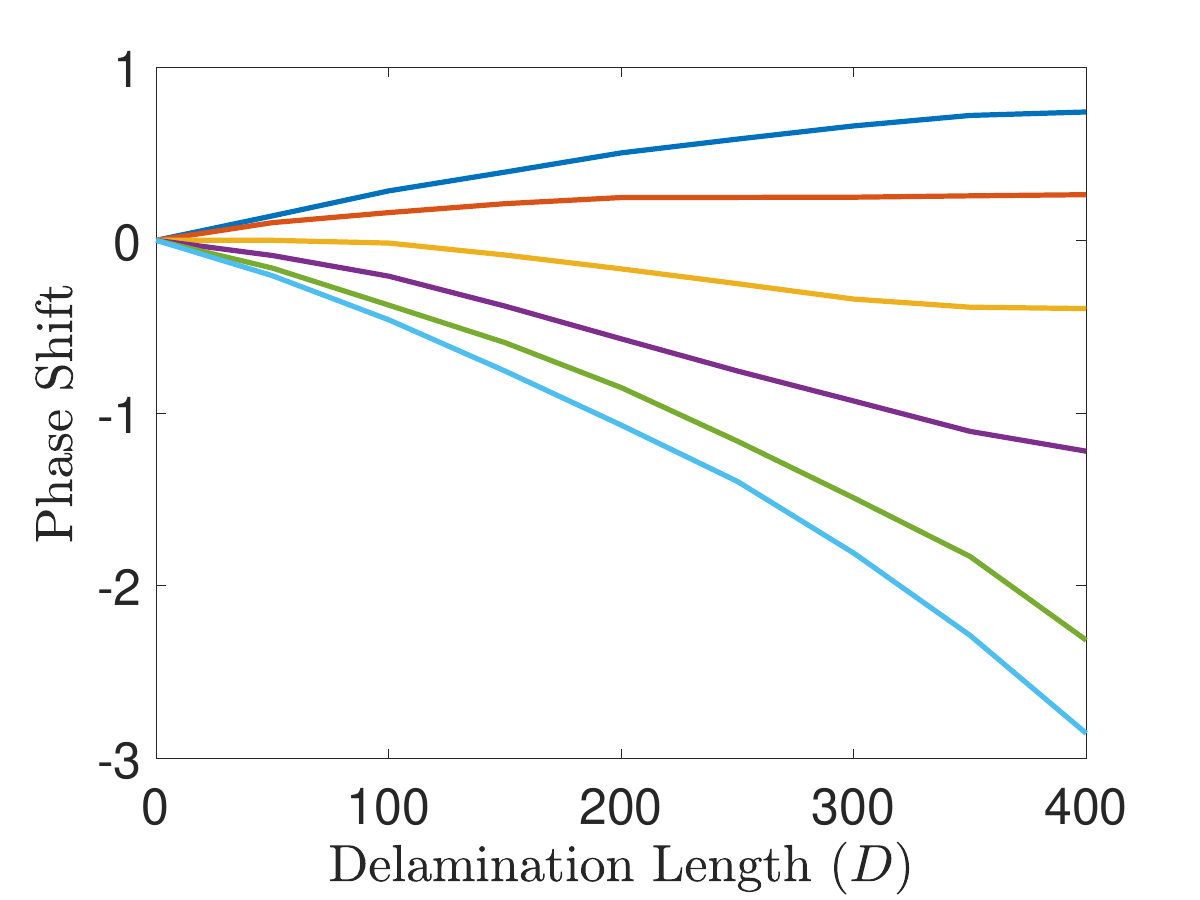}}
    \subfigure{\includegraphics[width=4.7cm]{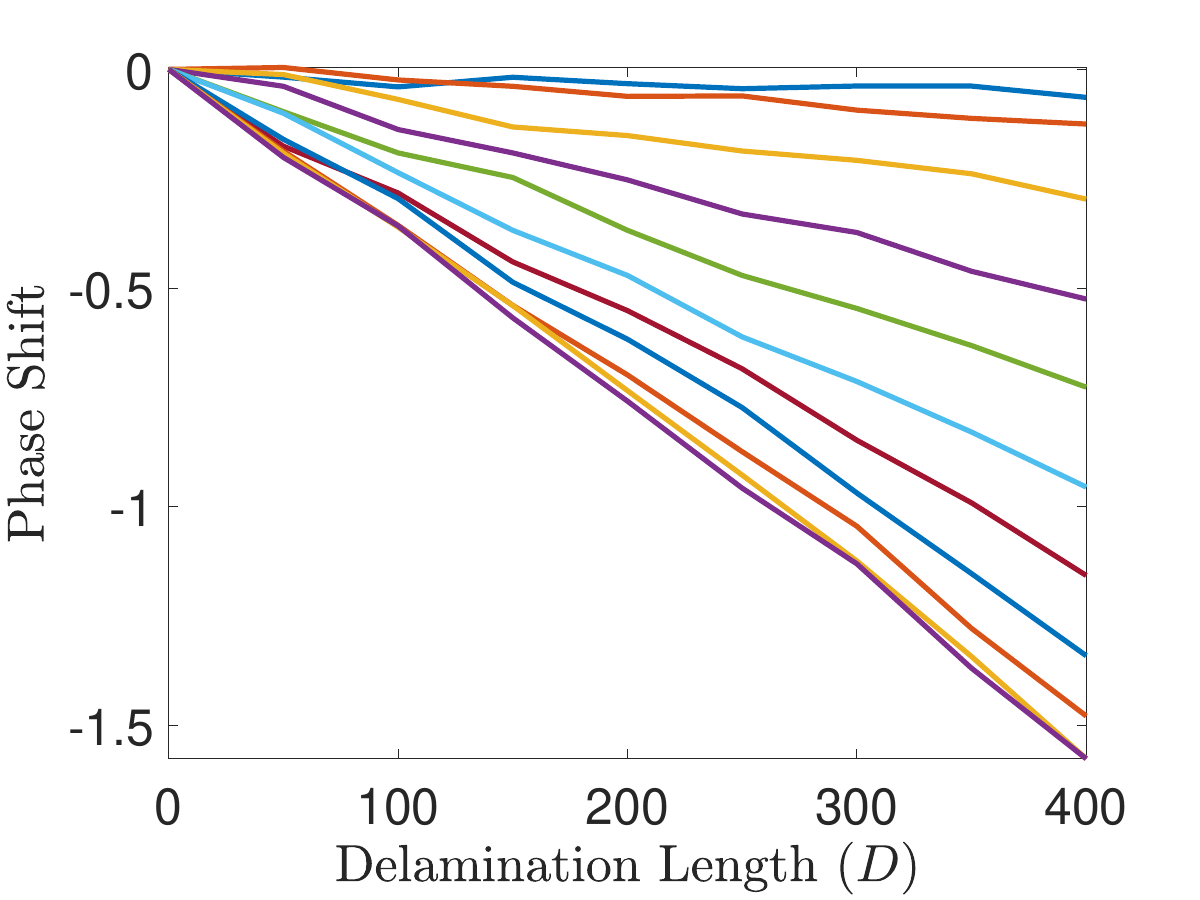}} \\
    \addtocounter{subfigure}{-2}
    \subfigure[$c = 1.25$]{\includegraphics[width=4.3cm]{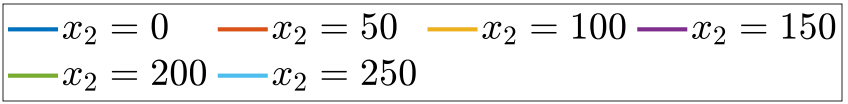}}
    \hspace{0.2cm}
    \subfigure[$c = 1.5$]{\includegraphics[width=4.3cm]{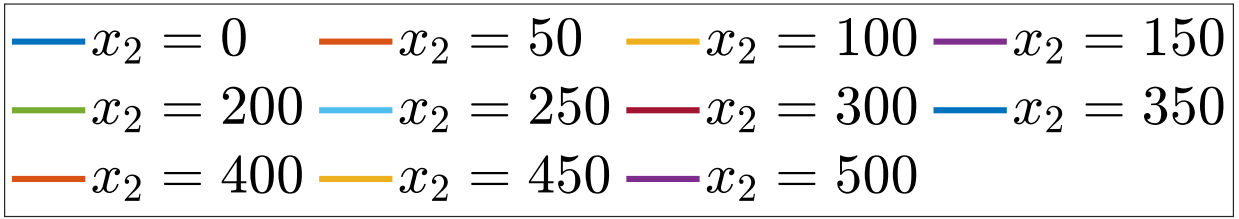}}
    \caption{`Fan' plots of the change in phase shift in the lower layer, with reference to the fully bonded case, compared against changes in delamination length, for (a) $c = 1.25$ and $x_2\in [0, 250]$, and (b) $c = 1.5$ and $x_2\in[0, 500]$. In both cases the delamination length $D$ is varied from 0 to 400.}
    \label{fig:FanLower}
\end{figure}

We can naturally obtain another measure for this structure by considering the lower section of the waveguide and performing the same analysis. We consider two cases, when $c = 1.25$ and $c = 1.5$, both of which are shown in Figure \ref{fig:FanLower}. These fans are similar to those for the upper layer, however for $c=1.5$, we observe more $x_2$ values prior to the reversal of gradient due to the decreased rate of evolution. Additionally, for $c=1.5$, we notice overlapping lines, attributed to the aforementioned slower evolution. This overlap becomes more prominent for larger values of $c$ and thus limits the applicability. However, we present a method to relax this limitation in Section \ref{sec:Reverse}.

We now outline a method to identify the position and size of the delamination using the results shown in Figures \ref{fig:FanUpper} and \ref{fig:FanLower}. Assume that we have performed experiments and detected phase shifts of $-1.5$ and $-2.01$ in the upper and lower layers, respectively.  Using the fans in Figure \ref{fig:FanUpper} we find intersections with the fan at $x_2 = 250,300,350$, with corresponding delamination lengths of $D=321,285,303$, respectively.  We then use the fans in Figure \ref{fig:FanLower} to find intersections with the fan at $x_2 = 200,250$ with corresponding delamination lengths of $D=296, 321$, respectively.  The only common value is for $x_2=250$ and $D=321$, thus we can identify a delaminated region from $x = 250$ to $x = 571$. This process could, in theory, provide multiple potential values.  However it could then be complemented with more conventional short wave methods at the identified locations to determine the correct value.

% Reverse propagation
\subsection{Reversing the sample}
\label{sec:Reverse}
As previously noted, for larger values of $c$ we observe a bunching of the branches in the fan due to less evolution (see Figure \ref{fig:FanLower}(b)). In this case, an observed phase shift will give a large number of possible values for $D$, while for even larger values of $c$ e.g. $c = 2$, the fan curves are almost superimposed, making it impossible to use the technique outlined in Section \ref{sec:Characterising}. Here we present an alternative approach.

\begin{figure}[h!] 
    \centering
    \subfigure{\includegraphics[width=4.7cm]{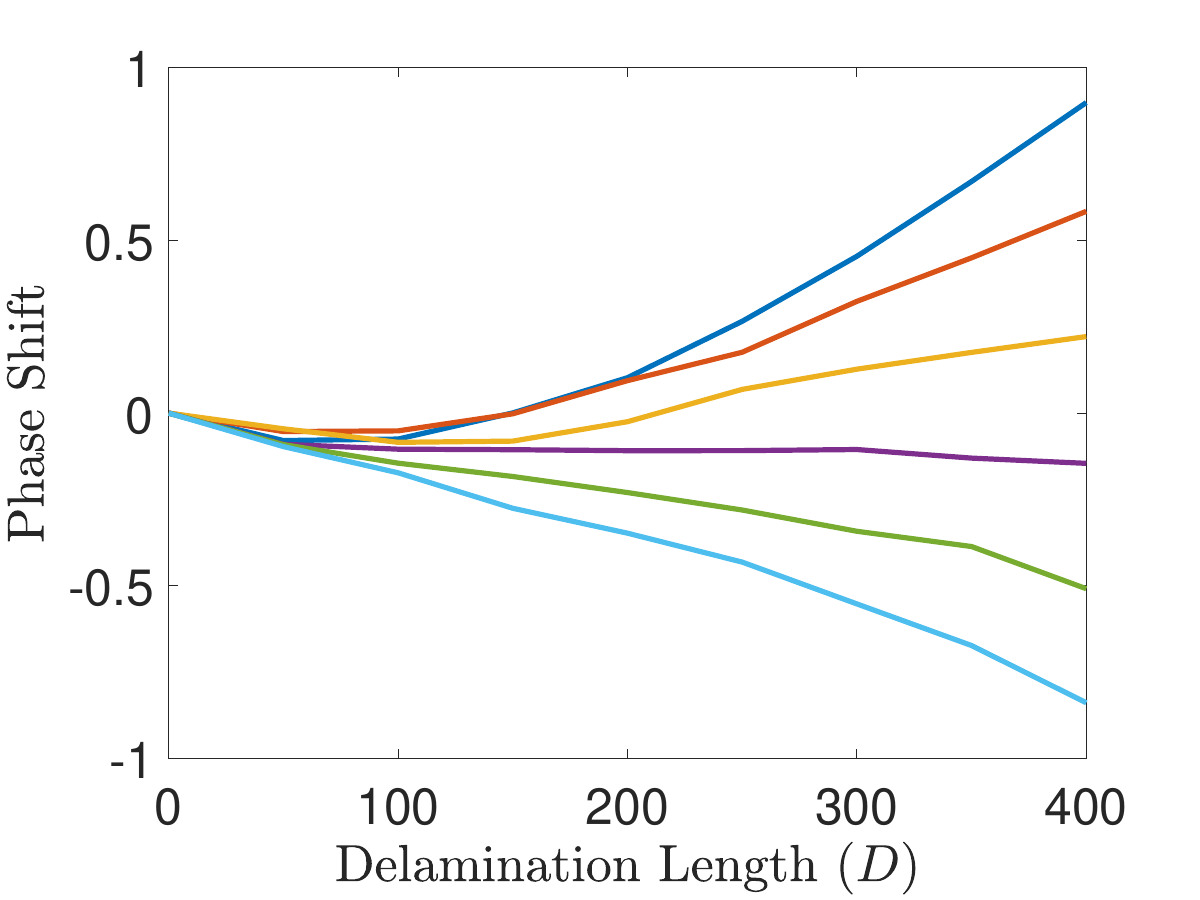}}\\
    \subfigure{\includegraphics[width=4.7cm]{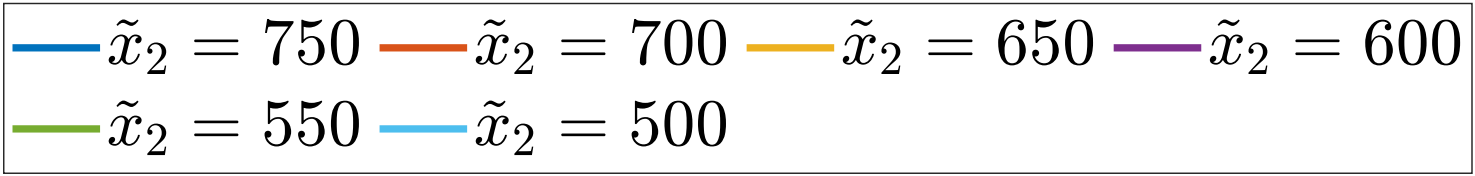}}
    \caption{`Fan' plot of the change in phase shift in the upper layer of the reversed sample, with reference to the fully bonded case, compared against changes in delamination length for $\tilde{x}_2\in [500, 750]$. The delamination length $D$ is varied from 0 to 400.}
    \label{fig:Backwards}
\end{figure}

For a given sample, we can send waves from left-to-right, or alternatively we can reverse the sample and send waves from right-to-left. In this reversed configuration, the initial homogeneous region is found between $\tilde{x}_4=2500$ and $\tilde{x}_3=2400 \:(= x_4)$. The interval from $\tilde{x}_3$ to $\tilde{x}_2\: (= x_3)$ constitutes the first soft bonded region, the delaminated region is from $\tilde{x}_2$ to $\tilde{x}_1 = \tilde{x}_2 - D \: (= x_2)$, while $\tilde{x}_0 = 0 \:(=x_1)$ is the end of the second soft bonded region. The fan generated in the upper layer for this case is presented in Figure \ref{fig:Backwards}.

The fan shape results here are similar to those found in Figure \ref{fig:FanUpper} for waves moving from left-to-right. To apply the results in Figure \ref{fig:Backwards} for identifying delamination, we assume that experiments were performed with incident waves in both directions and that the corresponding phase shifts were detected and recorded in the upper layer only where $c=1$. For example, assuming that we detect phase shifts of $-1.510$ for the left-to-right incident wave and $0.129$ for the right-to left incident wave, then we can check the fan intersections in Figures \ref{fig:FanUpper} and \ref{fig:Backwards}, respectively.  For the right-to-left case we find intersections at $\tilde{x}_2=750,700,650$ with corresponding delamination lengths of $D=207,221,300$. Hence the minimum possible coincident $x_2$ value for the left-to-right case is given by $x_2=650-D = 650-300=350$, since $x_2$ is taken at the left side of the delaminated region and $\tilde{x}_2$ at the right side. For the left-to-right case,  we therefore consider intersections in the range $x_2 \geqslant 350$ and obtain a unique intersection at $x_2 = 350$ (or $\tilde{x}_2=650$) with a delamination length of $300$. Hence, we can identify a delaminated region from $x = 350$ to $x = 650$. In practice, we may need to interpolate between curves to find the exact intersections; we only provide a proof of concept here.

% Conclusion
\section{CONCLUSIONS}
\label{sec:Conclusion}
In this paper we have considered the propagation of waves in a layered waveguide with soft bonding between the layers, including a delamination where the bonding has eroded. Previous studies have considered cases where the layers are identical with perfect bonding \cite{Khusnutdinova08, Khusnutdinova15}, and soft bonding with similar material properties \cite{Khusnutdinova17}. This study has examined the emergence of Ostrovsky wave packets in this system when the layers have distinct properties.

Direct numerical simulations can be performed for these problems \cite{Tranter19}, but we can learn more about the underlying physics through developing a semi-analytical approach based upon the use of several matched asymptotic multiple-scale expansions and averaging with respect to the fast space variable. This shows that the leading order waves are governed by Ostrovsky equations and thus Ostrovsky wave packets emerge. We have shown that the schemes are in good agreement around the main wave packet, with some divergence in the trailing radiation, which could be remedied by the inclusion of higher order corrections in the weakly-nonlinear solution, as shown for the corresponding initial-value problem in \cite{Khusnutdinova22}.

Our study has revealed that the presence of delamination leads to an amplitude decay and a phase shift in the Ostrovsky wave packet, giving a measure of the delamination. Using this phase shift, we constructed `fans' that represent the corresponding phase shift for different positions and lengths of delamination.  By performing experiments to detect these phase shifts, the intersections with the fans can be calculated for each layer and the common intersection gives the precise position and length of delamination. If the layers have significantly different properties, the fan in the lower layer is condensed and provides limited information, so the sample can be reversed and a second fan created for the waves propagating in the opposite direction through the upper waveguide. The intersections can then be compared for both fans to identify the coincident value(s).

This is a significant step forward in the field, as it is the first time to our knowledge that nonlinear waves have been shown to identify both the position and length of delamination, in contrast to previous studies \cite{Khusnutdinova15, Khusnutdinova17, Tamber22}. If multiple values are identified, linear wave methods could be used to supplement our results at the identified locations to determine the precise position and length.

Solitons have been observed in a layered PMMA waveguide with cyanoacrylate between the layers \cite{Dreiden08, Dreiden11, Dreiden12, Dreiden14} and radiating solitary waves have been observed in a two-layered PMMA bar with PCP (polychloroprene-rubber-based) adhesive \cite{Dreiden12}.  Our study motivates further laboratory experimentation with soft bonding and distinct material properties in each layer to confirm these estimates and provide a significant step towards a new technology that can be deployed in industrial settings.

% Bibliography
\bibliographystyle{elsarticle-num}
\bibliography{Research}
\end{document}